# Gene splice sites correlate with nucleosome positions

Simon Kogan and Edward N. Trifonov*

*Genome Diversity Center, Institute of Evolution, University of Haifa,*

*Mount Carmel, Haifa 31905, Israel*

## *Abstract*

Gene sequences in the vicinity of splice sites are found to possess dinucleotide periodicities, especially RR and YY, with the period close to the pitch of nucleosome DNA. This confirms previously reported finding about preferential positioning of splice junctions within the nucleosomes. The RR and YY dinucleotides oscillate counterphase, i.e., their respective preferred positions are shifted about half-period one from another, as it was observed earlier for AA and TT dinucleotides. Species specificity of nucleosome positioning DNA pattern is indicated by predominant use of the periodical GG(CC) dinucleotides in human and mouse genes, as opposed to predominant AA(TT) dinucleotides in Arabidopsis and C.elegans.

## *1. Introduction*

Nucleosomes, the building blocks of eukaryotic chromatin, play a key role in the organization of the overall chromatin structure and participate in regulated gene functioning [Anderson and Widom 2000; Ehrenhofer-Murray 2004]. Chromatin can also be involved in protection of functionally important sequence sites. In particular, the gene splicing sites appear to be protected by nucleosomes centered at the sites [Denisov, Shpigelman et al. 1997]. With large modern sequence databases at hand one can investigate this possibility in greater details.

Nucleosome DNA has been shown to display AA(TT) and, more generally, RR(YY) dinucleotide periodicity with the period close to the pitch of the DNA double helix [Trifonov and Sussman 1980; Mengeritsky and Trifonov 1983; Uberbacher, Harp et al. 1988; Kato, Onishi et al. 2003]. The dinucleotide distribution obtained by alignment of 4000 splice site sequences [Denisov, Shpigelman et al. 1997] did not show the periodicity, apparently, because of the small size of the data set. In this work we used much larger sequence data bases (40000 and 200000 splice site sequences). Clear periodicities are observed, with the amplitudes sufficient to estimate the period and the phase relationships between oscillating components.

## *2. Methods*

### **2.1. Exon-intron data base**

A comprehensive data base of protein-coding intron-containing nucleotide sequences [Saxonov, Daizadeh et al. 2000] was used in this work. Two data sets were utilized: a general one (both experimentally determined and algorithmically predicted genes from NCBI GenBank) containing

* Corresponding author.

Phone: 972-4-8288096

E-mail: trifonov@research.haifa.ac.il

about 50000 sequences and 200000 splice sites, and the data set of experimentally determined splice sites obtained by comparing genomic and mRNA sequences, with about 10000 sequences and 40000 splice sites. Most of this work is done with this latter data set, in order to avoid algorithmic biases and, thus, to increase reliability of results. In species-specific sequence analyses, the general data set was used.

## 2.2. Alignment by splice site

To obtain dinucleotide distributions in a neighborhood of a splice site, nucleotide sequences of the sets were aligned separately by exon-intron (EI) and intron-exon (IE) junctions. We utilized only those sequence segments that have at least 100 bases upstream and downstream from the splice site. The distributions in the interval [-100, 100] were calculated for each type of the junctions.

## 2.3. Window Fourier Transform

To estimate local periodicity of the sequence, Discrete-Continuous Fourier Transform [Oppenheim and Schafer 1975] was calculated in the moving window of specified size (the technique is further referred to as WFT). The window size was set to 42 nucleotides (approximately 4 nucleosome DNA periods). This size corresponds to estimated earlier periodical segments of the nucleosome DNA patterns [Boffelli, De Santis et al. 1991; Kato, Onishi et al. 2003; Cohanim, Yechezkel et al. 2004].

## 2.4. Sine best fit

To compare amplitudes of periodical components of the dinucleotide distributions and calculate respective phase shifts, sine function was fitted (by nonlinear least-squares minimization) to the distribution data in the specified window. The sine wave period was set equal to the nucleosome sequence DNA period 10.38 bases [Trifonov 1995].

The phases were calculated as a distance (in nucleotide units) between the splice site and the maximum of the fitting sine wave, by modulo of the sine period.

## *3. Results*

## 3.1. Dinucleotide periodicity in the vicinity of splice site

Inspection of all 32 distribution plots for 16 dinucleotides within ±100 bases from the EI and IE junctions shows that about 1/3 of the curves display apparent oscillations. It is especially obvious in case of purine-purine (RR) and pyrimidine-pyrimidine (YY) dinucleotides. The most conspicuous examples are shown in Fig. 1 (other distributions can be viewed at http://leah.haifa.ac.il/~skogan/splice-site-nucleosome/fig1a.htm ). Periodical segments of the curves, with the period equal approximately to 10 nucleotides are seen in the exon parts of the distributions. Some periodicity is detected as well in intron parts (not shown). To evaluate the observed periodicities in a more quantitative way, we applied WFT to the whole set of the distributions. Sum of WFT magnitudes for all 16 dinucleotide distributions as a function of period is presented in Fig. 2. Major peaks are seen in all WFT periodograms in the range of 9 – 11 bases in both EI and IE exons and EI intron as well. The average of the peak positions is $10.58 \pm 0.35$ that is close to known weight average value (10.38) of nucleosome sequence periodicity [Trifonov 1995; Ioshikhes, Bolshoy et al. 1996]. This value is used in further analyses.

To evaluate statistical significance of the oscillations, we generated 113 data sets of shuffled sequences, with preservation of overall dinucleotide compositions within given range, and processed them in the same way as the original data set. The calculated statistical significance (p-value) is 0.009, 0.062 and 0.044 for EI exon, IE exon and EI intron (Fig. 2) respectively. For EI exon, none of random (shuffled) data sets gave higher peak than the original data set. That is, the above result is actually a p-value's upper boundary. Noteworthy that the whole set of the dinucleotide distributions was used for the above estimates of the effect, including many of the distributions that contributed only noise.

To further investigate the periodical signal in the vicinity of the splice sites, we fitted sine wave (with fixed period 10.38) to dinucleotide distributions. The fitting sine window size was set to 42 nucleotides for the same reason as a window size for WFT procedure. Hence, only four peaks in each plot of the Fig. 1 are marked. The window center position was fixed for certainty at 40 nucleotides from splice junction, to span altogether 122 bases that is close to the estimated nucleosome core DNA size, 125 nucleotides [Kato, Onishi et al. 2003, and references therein].

The window parameters, though, are not critical. The sine fitting procedure in the windows positioned 45 and 50 nucleotides from the splice junctions give very similar results (data not shown, see also Fig. 2).

### 3.2. Phase shift

In the Table 1 the amplitudes and phases of oscillating components of all RR and YY distributions are presented. The data is sorted according to fitting sine amplitudes, separately for EI and IE exons. Only 5 of them are significant at the 5% level (7 are significant at 10% level).

Remarkably, the phases of the oscillating dinucleotides cluster in two separate regions, centered at $0.9\pm0.7$ for YY and $7.0\pm0.7$ for RR dinucleotides, irrespective of their upstream or downstream positions relative to the splice junctions. The phase shift between RR and YY is 6.1 bases, close to a half-period, and similar to the shift between AA and TT dinucleotides of nucleosome positioning DNA pattern in [Ioshikhes, Bolshoy et al. 1996].

Important to note that the phases of the oscillations are apparently identical for exons of both EI and IE junctions. That would be expected in case when the splice junctions within the nucleosomes are strictly phased with the nucleosome sequence periodicity and, thus, specifically oriented relative to the nucleosome surface (see Discussion).

Some of RY and YR dinucleotides display 10 – 11 base periodicity as well (see Fig. 1). However, they do not show as consistent stability of phases as RR(YY) dinucleotides do. Their amplitudes are generally lower (see also below). More massive data would be required to relate these oscillations to nucleosome positioning DNA pattern.

### 3.3. Species specificity of RR(YY) oscillations

As the data suggest the nucleosome positioning sequence pattern appears to involve not only AA and TT dinucleotides which have been considered as dominant factor in earlier studies, but other RR and YY dinucleotides as well. In our recent work with human nucleosome DNA sequences [Kato, Onishi et al. 2003] the AA(TT) dinucleotides failed to show expected dominance, while RR(YY) dinucleotides as a group did display the periodicity. This points to possible species-specificity of the nucleosome sequence pattern. To explore this matter, we performed the same analysis as above on species-specific subsets of the original data set. The results are presented in the Table 2 where the dinucleotide periodicities around splice junctions are compared for human, mouse, Arabidopsis and C. elegans. Human and mouse sequences show very clear domination of GG and CC dinucleotides. Neither AA nor TT appear among the eight strongest periodical contributions in human sequences. On the other hand, neither GG nor CC appear in the list of the strongest for C. elegans. RR and YY dinucleotides of one or another type are major contributors in all four cases.

## *4. Discussion*

Utilizing the large data set of gene sequences, we were able to observe RR(YY) periodical pattern (with the period close to the pitch of nucleosome DNA) in the vicinity of gene splice sites. This supports previously reported connection between nucleosome and splice site positioning [Denisov, Shpigelman et al. 1997]. Our estimation of a phase shift between RR and YY periodical components of the presumed nucleosome positioning DNA pattern is approximately 6 nucleotides that is in agreement with the earlier estimate for the AA(TT) nucleosome sequence pattern [Ioshikhes, Bolshoy et al. 1996]. This behavior of RR and YY dinucleotides around splice junctions is, thus, fully consistent both with the nucleosome sequence pattern and with the connection between splice sites and the nucleosome positions.

The periodical behavior of RR(YY) dinucleotides in general, rather than just AA(TT) dinucleotides, has been originally observed in [Mengeritsky and Trifonov 1983] on a small ensemble of chromatin DNA sequences. There as well the RR(YY) dinucleotides other than AA(TT) behaved in a similar way – the same periodicity and the same phase. Nucleosome sequence database of [Ioshikhes, Bolshoy et al. 1996] apart from AA(TT) component also contained periodical GG(CC) dinucleotides [Bolshoy 1995]. In our recent study [Kato, Onishi et al. 2003] with human nucleosome sequences we, indeed, observed the oscillation of RR(YY) dinucleotides, while the amplitude of AA(TT) oscillation was negligible. It is fully consistent with the observation in this work (see Table 2) that suggests that the major oscillatory contribution in human genes is the one of GG and CC.

Our analysis reveals species-specificity of nucleosome DNA pattern. The data indicate that while the nucleosome sequences of C. elegans are AA(TT) periodical, as well as of SV40 virus in an early work [Trifonov and Sussman 1980] and of S. cerevisiae , in more recent studies [Herzel, Weiss et al. 1999; Tomita, Wada et al. 1999; Cohanim, Yechezkel et al. 2004], other organisms, like human and mouse, display nucleosome sequence periodicity in form of oscillating GG and CC, as well as AG(CT) dinucleotides in smaller proportions, see also [Kato, Onishi et al. 2003].

From the Table 2 it follows that neither GG(CC) periodical nor AA(TT) periodical genomes display any appreciable periodicity of TA dinucleotide. Periodicity of this dinucleotide is documented as characteristic of selected most stable nucleosomes [Thastrom, Lowary et al. 1999]. In our work the emphasis has been always on the patterns that are characteristic of bulk chromatin. It appears, thus, that the most common pattern for the typical natural nucleosomes would be the counterphase oscillation of any or all of RR and YY dinucleotides with the period about 10.4 bases, and with the same phase for all RR (and YY, respectively) dinucleotides.

Of special interest is the positioning of the splice junctions relative to the surface of the nucleosome. In the earlier work on the positioning [Denisov, Shpigelman et al. 1997] the splice junctions EI and IE were found both located 3-5 bases downstream from the central point of the nucleosome DNA. In that work the nucleosomes have been mapped by means of the AA(TT) nucleosome pattern [Ioshikhes, Bolshoy et al. 1996], where the phase of AA oscillation is 8±1 bases downstream of the nucleosome DNA center – on the interface with the histone octamer. In the current work the phase for AA (and any RR as well) is 7±0.7 bases counting from the junction. That is, according to the above numbers, the obligatory GT and AG dinucleotides at the junctions are about 1-2 bases downstream from the nucleosome DNA central point. Despite the poor accuracy, both estimates of the positions of the junctions indicate that GT and AG dinucleotides at the intron ends are close to external positions in the DNA molecule on the surface of the nucleosome. The most external point of the nucleosome DNA is 2.6 bases downstream from the nucleosome DNA center, in the 5'-3' sequence direction. The bases of DNA molecule at this phase position and close to it are farthest away from the surface of the histone octamer. This, of course, means that in the template strand of DNA, the AC and CT dinucleotides, complementary to the canonical intron ends, are both close to the histone surface. Such safe placement one would expect *a priori* if the fidelity of copying the splice junctions is to be maximized. This gives an important rationale for preferential positioning of the splice junctions close to the nucleosome DNA center, on outward surface of the nucleosome DNA. The sequence periodicities observed in this work, apparently, correspond to the fraction of the nucleosomes that, indeed, are involved in the protection of the splice junctions.

## *Figure and table legends*

### Table 1 legend

Amplitudes and phases of fitting sine waves for all RR and YY dinucleotide distributions in the vicinity of exon-intron and intron-exon splice junctions, calculated from verified mRNA data set. The phases were calculated as a distance (in nucleotide units) between the splice site and the maximum of the fitting sine wave, by modulo of the sine period.

### Table 2 legend

Weighted amplitudes of dinucleotide periodicities (arranged in descending order) around splice junctions of Human, Mouse, Arabidopsis and C. elegans genes calculated from general data set. Only topmost eight amplitudes are presented in each case. For the purpose of comparison the weights are calculated as dinucleotide fitting sine amplitude divided by the sum of all 32 amplitudes, for each species separately.

### Figure 1 legend

Fig. 1. The most conspicuous examples of dinucleotide distributions in the vicinity of exon-intron and intron-exon splice junctions showing the 10-11 base periodicity. The data set verified by mRNA was used for the calculation. X-axis - dinucleotide position counted from the splice junction. Y-axis -dinucleotide count. Dots correspond to peaks of the fitting sine wave with the period 10.38. The fitting sine window is centered at 40 nucleotides from splice junctions.

### Figure 2 legend

Fig. 2. Sum of WFT amplitudes of all 16 dinucleotide distributions as a function of period. X-axis - period in bases. Y-axis - WFT amplitudes for windows positioned 40, 45 and 50 nucleotides from their centers to splice site (lines —, ---- and -.-.- respectively). Vertical lines correspond to period 10.38. Average position of the maxima for both EI and IE exons and EI intron (9 plots) is $10.58 \pm 0.35$.

Table 1. Sine fit parameters for RR and YY dinucleotide distributions

| Junction | Sequence region | Dinucleotide | Amplitude | P-value | Phase |
|---|---|---|---|---|---|
| EI | exon | CC | 84.96 | 0.027 | 0.92 |
| EI | exon | GG | 65.76 | 0.035 | 7.76 |
| EI | exon | CT | 38.67 | 0.212 | 1.96 |
| EI | exon | TC | 36.39 | 0.053 | 0.17 |
| EI | exon | AG | 28.87 | 0.265 | 6.94 |
| EI | exon | TT | 28.45 | 0.239 | 2.59 |
| EI | exon | AA | 25.63 | 0.504 | 5.80 |
| EI | exon | GA | 12.31 | 0.770 | 9.89 |
|  |  |  |  |  |  |
| IE | exon | AA | 68.82 | 0.009 | 6.48 |
| IE | exon | AG | 47.58 | 0.009 | 6.69 |
| IE | exon | CT | 46.45 | 0.088 | 0.68 |
| IE | exon | TT | 34.51 | 0.071 | 1.88 |
| IE | exon | CC | 27.99 | 0.788 | 0.37 |
| IE | exon | TC | 18.50 | 0.496 | 0.69 |
| IE | exon | GA | 11.87 | 0.912 | 0.50 |
| IE | exon | GG | 9.84 | 0.876 | 5.67 |

Table 2. Amplitudes of dinucleotide distribution periodical components in four species

| Human | | Mouse | | Arabidopsis | | C.elegans | |
|---|---|---|---|---|---|---|---|
| IE exon GG | 0.094 | EI exon GG | 0.091 | IE exon AA | 0.064 | EI exon AA | 0.082 |
| EI exon GG | 0.084 | IE exon CC | 0.085 | IE exon CA | 0.059 | IE exon AC | 0.059 |
| EI exon CC | 0.081 | IE exon GG | 0.064 | IE exon GA | 0.059 | IE exon CA | 0.057 |
| EI exon TG | 0.058 | EI exon TT | 0.050 | EI exon GG | 0.051 | IE exon AT | 0.052 |
| IE exon TG | 0.053 | EI exon CC | 0.046 | IE exon AT | 0.050 | IE exon TT | 0.051 |
| IE exon CC | 0.053 | IE exon AG | 0.045 | IE exon GG | 0.048 | IE exon CT | 0.042 |
| EI exon AC | 0.050 | IE exon TG | 0.043 | IE exon TG | 0.047 | EI exon AG | 0.040 |
| EI exon CT | 0.050 | EI exon GT | 0.040 | EI exon TT | 0.042 | EI exon CA | 0.039 |

**Fig. 1**

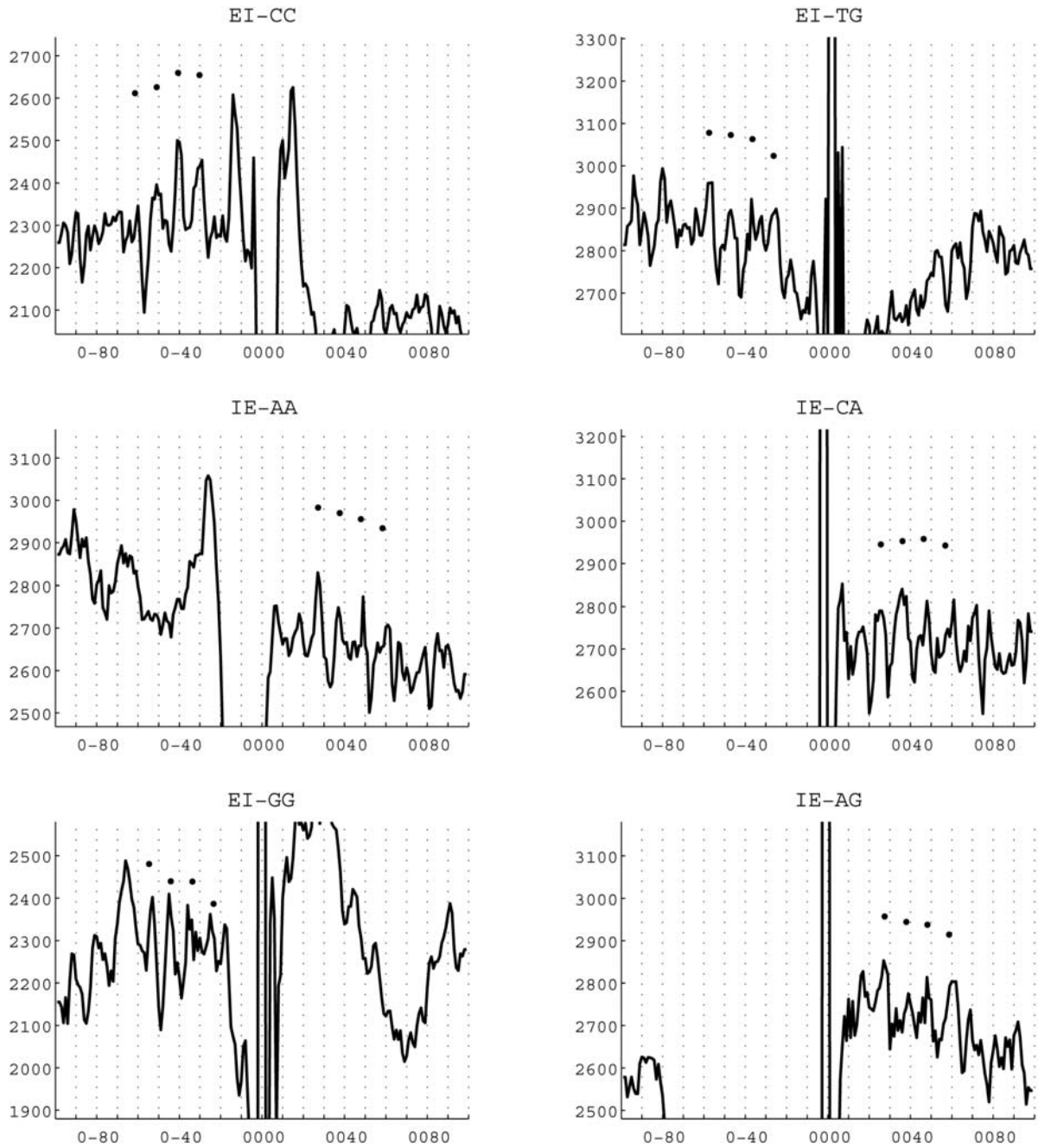

**Fig. 2**

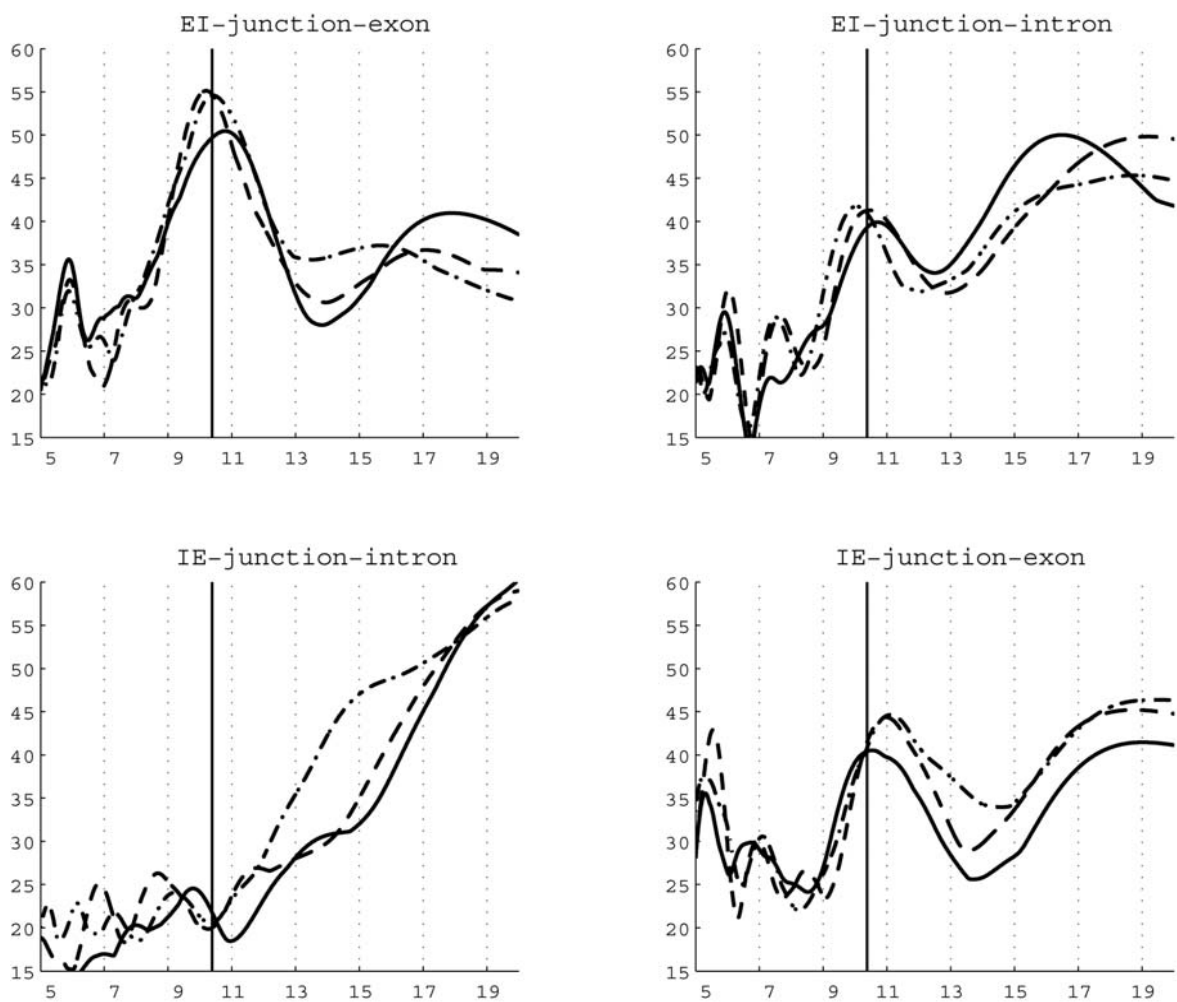